\documentclass[apj]{emulateapj}
\usepackage{natbib}
\newcommand\cubi{${\it c}_{\rm U,B,I}$}

\shorttitle{The Carina Project}
\shortauthors{Monelli et al.}

\begin{document}

\title{The Carina project VII: towards the breaking of the age-metallicity degeneracy 
of red giant branch stars using the ${\it C}_{\rm U,B,I}$\ index}

\author{M.\, Monelli\altaffilmark{1,2},
A.\, P.\, Milone\altaffilmark{1,2,3},
M.\, Fabrizio\altaffilmark{4},
G.\, Bono\altaffilmark{5},
P.\,B.\, Stetson\altaffilmark{6},
A.\,R.\, Walker\altaffilmark{7},
S.\, Cassisi\altaffilmark{4},
C.\, Gallart\altaffilmark{1,2},
M.\, Nonino\altaffilmark{8},
A.\, Aparicio\altaffilmark{1,2},
R.\, Buonanno\altaffilmark{5,4},
M.\, Dall'Ora\altaffilmark{9},
I.\, Ferraro\altaffilmark{10},
G.\, Iannicola\altaffilmark{10},
L.\, Pulone\altaffilmark{10},
F.\, Th{\'e}venin\altaffilmark{11}
}

\altaffiltext{1}{Instituto de Astrof\'{i}sica de Canarias, Calle Via Lactea s/n, 38205 La Laguna, Tenerife, Spain}
\altaffiltext{2}{Departamento de Astrof\'{i}sica, Universidad de La Laguna, 38200 La Laguna, Tenerife, Spain}
\altaffiltext{3}{Research School of Astronomy and Astrophysics, The Australian National University, Cotter Road, Weston, ACT, 2611,Australia}
\altaffiltext{4}{Istituto Nazionale di Astrofisica-Osservatorio Astronomico Collurania, Via M. Maggini, 64100 Teramo, Italy.}
\altaffiltext{5}{Dipartimento di Fisica, Universit\'{a} di Roma Tor Vergata, Via della Ricerca Scientifica 1, 00133 Rome, Italy}
\altaffiltext{6}{Dominion Astrophysical Observatory, NRC-Herzberg, 5071 West Saanich Road, Victoria, BC V9E 2E7, Canada}
\altaffiltext{7}{Cerro Tololo Inter-American Observatory, National Optical Astronomy Observatory, Casilla 603, La Serena, Chile.}
\altaffiltext{8}{Istituto Nazionale di Astrofisica-Osservatorio Astronomico di Trieste, Via G.B. Tiepolo 11, 40131 Trieste, Italy.}
\altaffiltext{9}{INAF- Osservatorio Astronomico di Capodimonte, Salita Moiariello 16, I - 80131 Napoli, Italy.}
\altaffiltext{10}{Istituto Nazionale di Astrofisica-Osservatorio Astronomico di Roma, Via Frascati 33, Monte Porzio Catone, Rome, Italy.}
\altaffiltext{11}{Universit\'{e} de Nice Sophia-antipolis, CNRS, Observatoire de la C\^{o}te d'Azur, Laboratoire Lagrange, BP 4229,06304 Nice, France.}

\begin{abstract}

We present an analysis of photometric and spectroscopic data of the Carina
dSph galaxy, testing a new approach similar to that used to disentangle
multiple populations  in Galactic globular clusters (GCs).
We show that a
proper colour combination is able to separate a significant  fraction of the
red giant branch (RGB) of the two main Carina populations  (the old one,
$\sim$12 Gyr, and the intermediate--age one, 4$-$8 Gyr).  In particular,  the
\cubi\ = $(U-B)-(B-I)$ pseudo-colour allows us to follow  the RGB of both
populations along a relevant portion of the RGB. We find that the  oldest
stars have more negative \cubi\ pseudo-colour than intermediate--age ones. We
correlate  the pseudo-colour of RGB stars with their chemical properties,
finding a significant trend between the iron content and the \cubi. Stars
belonging to the old population are systematically more metal poor
([Fe/H]$=-2.32\pm0.08$ dex) than the intermediate--age ones 
([Fe/H]$=-1.82\pm0.03$ dex).
This gives solid evidence on the chemical
evolution history of this galaxy, and we have a new diagnostic that can allow
us to break the age-metallicity  degeneracy of H-burning advanced evolutionary
phases. We compared the distribution of stars in the \cubi\ plane with
theoretical isochrones, finding that no satisfactory agreement can be reached
with models developed in a theoretical framework based on standard heavy
element distributions. Finally, we discuss possible systematic differences
when compared with multiple populations in GCs.

\end{abstract}


\section{Introduction}\label{sec:intro}

%
\begin{figure*}[t]
\centering
\includegraphics[width=16truecm, height=11truecm]{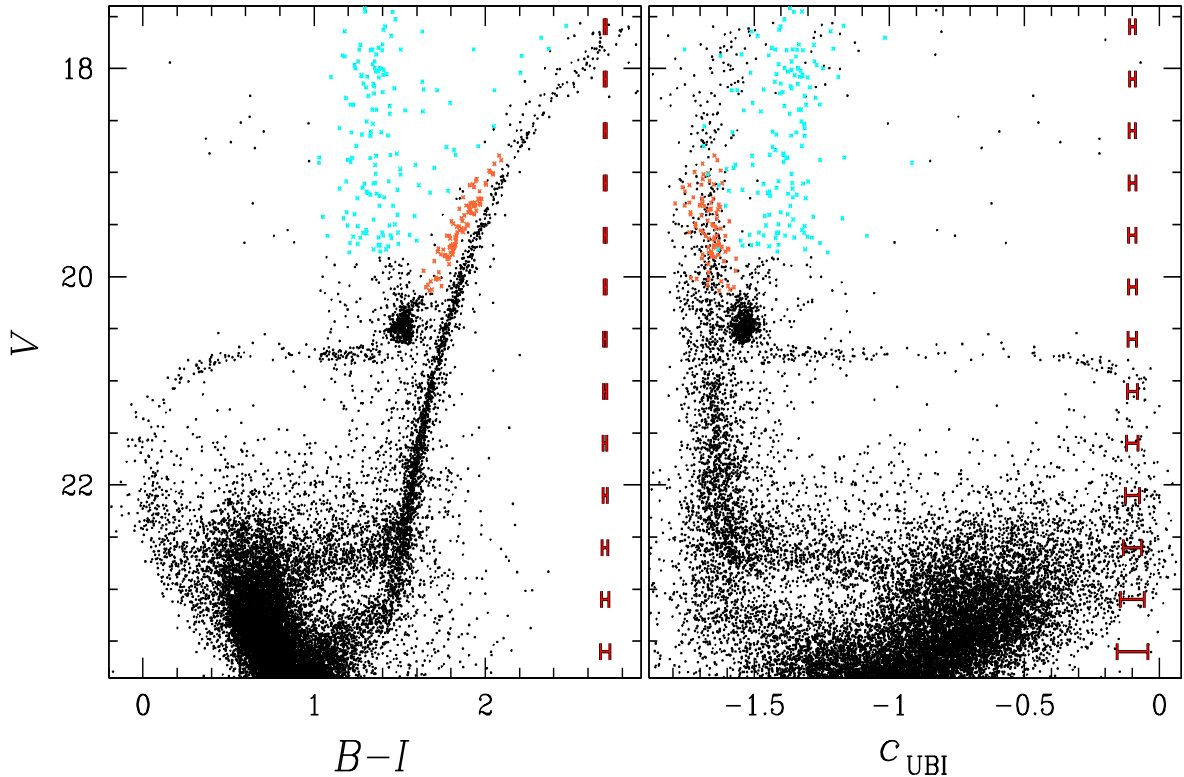}
\caption{{\em Left -} ($V$, $B-I$) CMD of Carina stars, based on the photometry
presented in \citet{bono10}. Bright field stars are highlighted in cyan, while
orange symbols show the Carina AGB stars;
{\em Right -} ($V$, \cubi) plane, where the RGBs of the old and 
intermediate--age populations largely split.  \label{fig:cubi}}
\end{figure*}
%

%
\begin{figure*}[t]
\centering
\includegraphics[width=16truecm, height=7.37truecm]{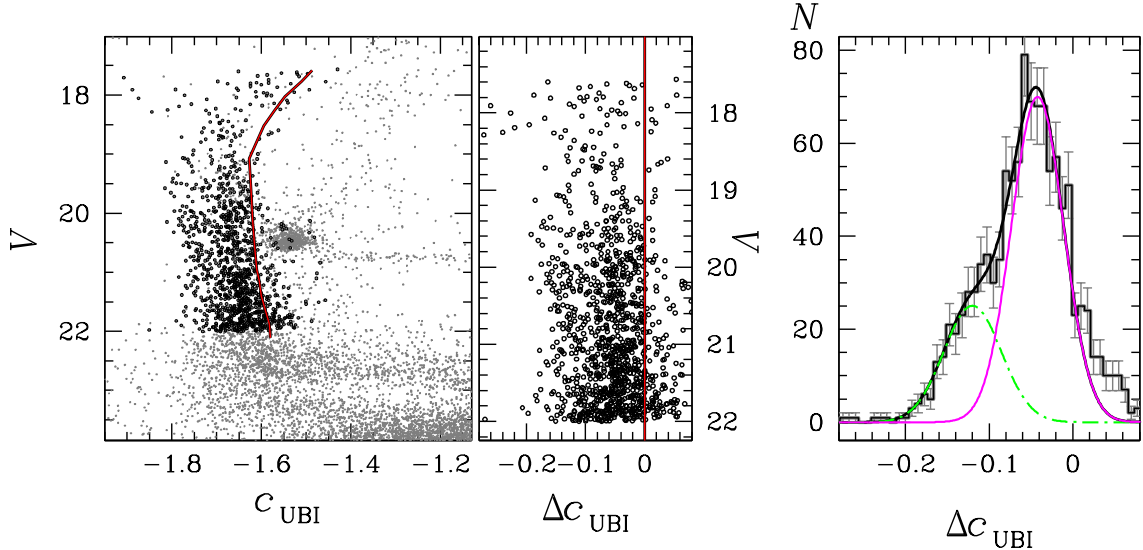}
\caption{ {\em Left)} The sample of RGB stars adopted to identify substructures 
in the RGB of Carina are shown as black open circles (1134 stars). The red 
line shows our adopted right envelope of the sequence. {\em Center -}
the Carina RGBs, rectified to the red line. For each star we display the 
$\Delta$\cubi\ index, the colour difference with respect to the 
reference red line. {\em Right -} Histogram of $\Delta$\cubi, with 
gaussian kernels superimposed. The overall distribution can be
well represented by two gaussian components.  \label{fig:cubi2}}
\end{figure*}
%

%
\begin{figure*}[t]
\centering
\includegraphics[width=16truecm, height=7.37truecm]{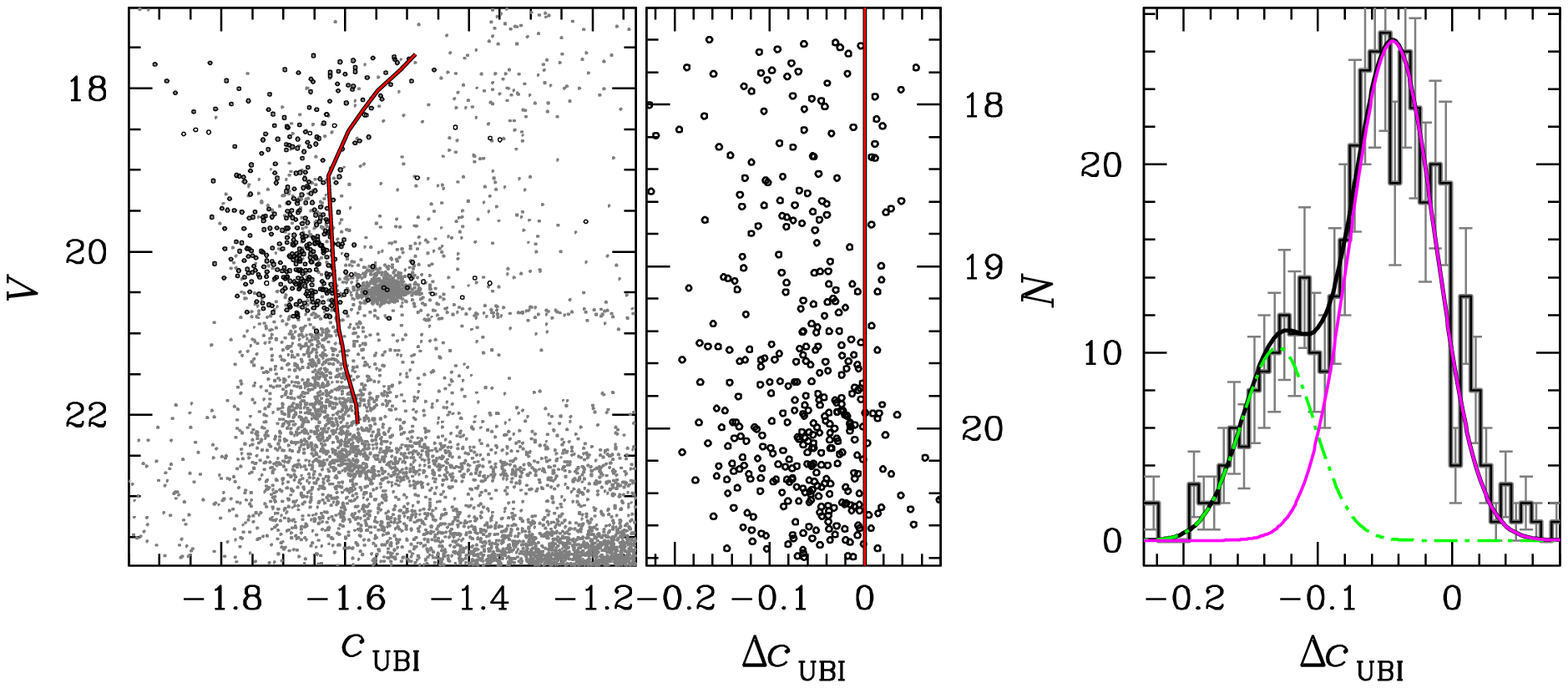}
\caption{Same as Figure \ref{fig:cubi2}, but in this case the open circles 
in the left panel mark
413 RGB stars with radial velocity consistent with that of Carina, selected
from the sample of \citet{fabrizio11}. In the right panel, the presence of
a double peak in the $\Delta$\cubi\  distribution is confirmed.  \label{fig:kinem}}
\end{figure*}
%

The classical question of how galaxies in general formed and evolved is one
that can be illuminated by the study of galaxies in the Local Group (LG).
Particularly the numerous dwarf galaxies that are thought in some respects
(mass, low metals) to be similar to the building blocks that have and are
still being accreted by more massive galaxies such as our own Galaxy and M31.
The characterization of the resolved stellar populations in terms of age and
chemical content of dwarf galaxies in the LG has been the target of large 
observational efforts in the last decades \citep[e.g.][]{gallart96a,gallagher98,held99,
aparicio01,bellazzini02,carrera02,dolphin01,skillman03,momany05,mcconnachie06,
cole07,rizzi07,walker09, martin09,monelli10a,monelli10b,sanna10,hidalgo11,
battaglia12a,dallora12,deboer12a,okamoto12}.

Until recently, dwarf galaxies were thought to be fundamentally different
from globular clusters (GCs). However, the recognition that many, if not all,
GCs contain multiple populations \citep{kraft92,cannon98,grundahl99, bedin04,
marino08,milone08, yong08,lee09,piotto12,monelli13} and that some of the
most massive clusters may be the remnants of dissipated dwarf galaxies
\citep{bekki07a} has blurred the differences somewhat. When considering
the build-up of the halo of our own Galaxy, its chemical evolution is likely
to be intimately intertwined with that of the GCs and accreting dwarf
galaxies. The first high resolution spectra for a handful of stars in
nearby dSphs \citep{shetrone01,shetrone03,tolstoy03} disclosed
a systematic difference in the chemical enrichment law compared to the
halo stars. This implies that the building blocks of the halo can not
resemble the {\itshape present day} Galaxy satellites, unless the merging
events occurred very early on, before the internal chemical evolution driven
by Supernova enrichment took place. On the other hand, it has recently been
suggested that GCs might be responsible for a significant fraction of the 
halo stars observed today \citep{dercole08}. In this scenario, the GCs 
formed two or more generations of stars, on a short time scale, of the
order of few hundred Myr, and the first Supernovae would be responsible
for a significant mass loss from the clusters, causing the evaporation 
of a large 
fraction of the first generation stars; these would build up part of
the Galaxy halo. Hence, a detailed knowledge of the chemical evolution of nearby
dwarfs and a comparison with GCs can give important insight on the
progenitors of our own Galaxy.

Ideally, we would like to collect high resolution spectroscopy for large samples of
stars to faint magnitudes in conjunction with accurate photometry. However
for LG galaxies the former is technically at present beyond our means and
only typically the brightest stars in the closer galaxies have such
data. Deep photometry alone suffers from the long-standing problem of the 
photometric colour-magnitude diagram (CMD) having an age-metallicity
degeneracy that affects the advanced evolutionary phases such as the 
red giant branch (RGB), where sequences of different age-metallicity 
combinations occupy the same region of the CMD and therefore can not
be easily disentangled. 

Interestingly, recent results revealed that the different 
stellar populations in old GCs can be easily isolated along the whole CMD, 
from the main sequence (MS), up to the subgiant branch (SGB), RGB, and even the 
horizontal branch by using an appropriate combination of broad band filters 
\citep{marino08,sbordone11,milone12e}. 
Monelli et al.\ (2013)\footnote{http://www.iac.es/project/sumo} showed that the
\cubi\ = $(U-B)-(B-I)$ index is a powerful tool to identify multiple stellar 
sequences in the RGB of old GCs, and that the \cubi\ pseudo-colour of RGB 
stars correlates with the chemical abundances of light elements. 

%
\begin{figure*}
\centering
\includegraphics[]{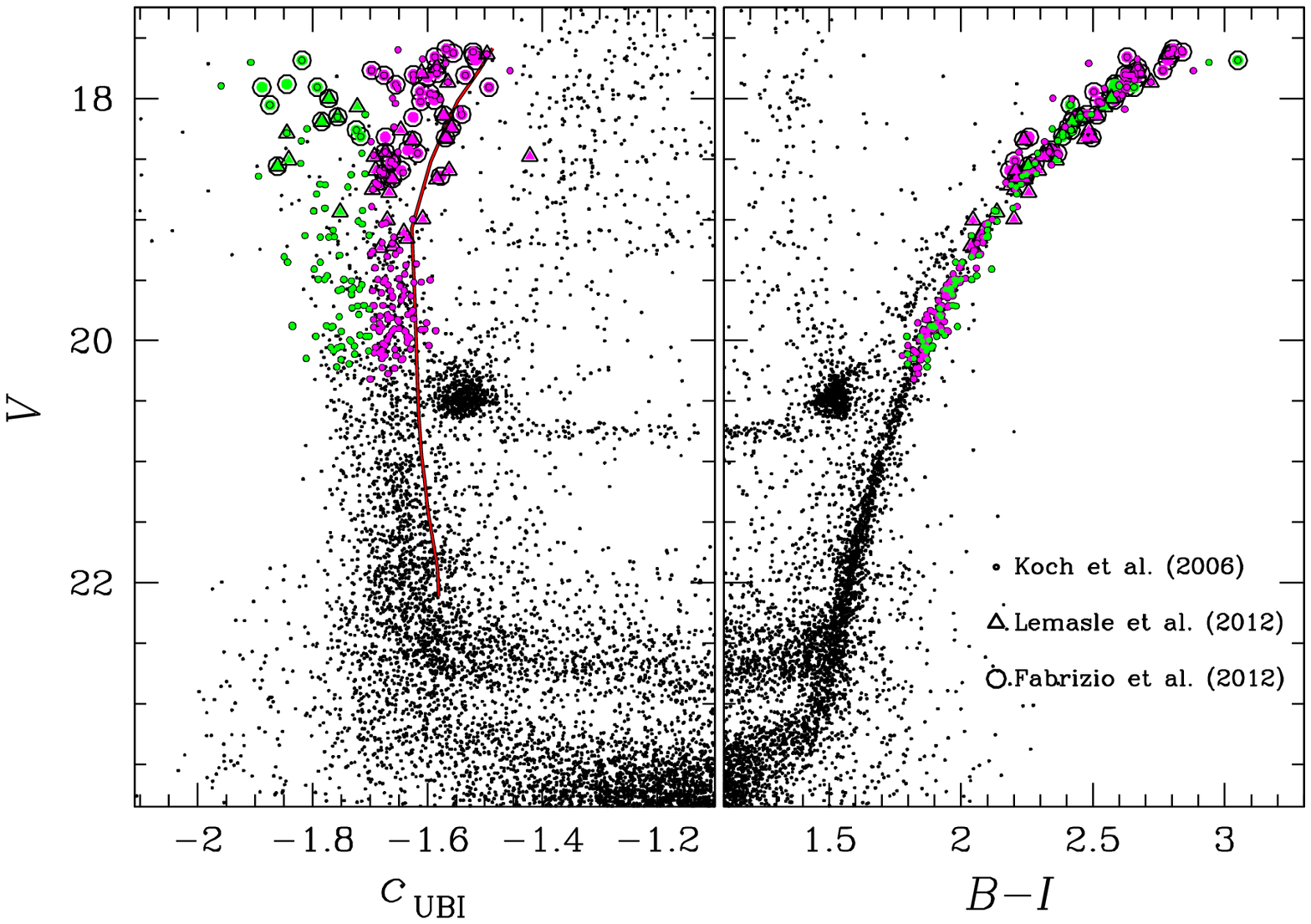}
\caption{{\em Left -} ($V$, \cubi) pseudo-CMD with superimposed the stars for
which  spectroscopic measurements are available. Large symbols show stars
with available Fe abundances from high-resolution spectra (dots: 
\citealt{fabrizio12}; triangles: \citealt{lemasle12}). Small coloured
dots mark the 212 stars selected from \citet{koch06}, for which the Fe abundance
was derived from CaT measurements. Green and magenta symbols indicate stars with
\cubi $< -1.70$ mag and \cubi $> -1.70$ mag, respectively. {\em Right -} ($V$,
$B-I$) CMD showing the complete degeneracy of spectroscopic targets on the RGB,
using the same colour coding.
\label{fig:cmd_target}}
\end{figure*}
%

In this paper we test the same approach successfully used for GCs for 
the first time in a dwarf galaxy. Since the established nomenclature can 
be somehow confusing, it has to be clearly stated that when we refer the 
stellar populations in a dwarf galaxy, in particular the dSph satellites
of the Milky Way, these have substantially different 
properties from multiple populations in GCs. First, the typical main star 
formation event in a dwarf galaxy occurred on time scales of the order of 
few Gyr, significantly longer than the age difference between two generations
of stars in a GC (few hundred Myr at most). Moreover, the chemical evolution 
is different. In particular, the large spread in Fe observed in dwarf galaxies
has no counterpart in a typical GC. In these systems, the internal chemical 
enrichment affected only the light elements involved in the high temperature 
H burning processes, while typically no spread in Fe, is found. 

In this paper we focus on the properties of stellar population in the Carina 
dSph, for which an extensive database of both photometric 
\citep{bono10,battaglia12b} and spectroscopic \citep{koch06,helmi06,
fabrizio11,fabrizio12,venn12,lemasle12} data is available. Nevertheless, 
we still lack a definitive understanding of its bursty star formation history 
\citep{smeckerhane96,io, stetson11}, unique in the Local Group, and of the 
chemical enrichment history of its different stellar populations. In fact, accurate 
analysis of high-quality photometry revealed that Carina experienced two 
main events of star formation, separated by a quiescent phase lasting few
Gyr, which produced two well separated SGBs. One corresponds to
an old population ($>$10 Gyr) containing approximately one third of the
total number of stars, while the second belongs to an intermediate--age
population made of $\sim$4-8 Gyr-old stars \citep{smeckerhane96,io}. The
two SGBs merge in a narrow RGB, making it difficult to associate the 
individual iron abundances with the discrete SGB and main sequence
\citep{koch06,fabrizio12,venn12}, observed in the CMD. This represents a major 
challenge to understanding the complex star formation history and the chemical 
evolution of the Carina dSph.

The paper is organized as follows. In section \ref{sec:data} we present the \cubi\
diagram for Carina, and the correlation between the \cubi\ pseudo-colour and the Fe
content of RGB stars. In section \ref{sec:discussion} we discuss the properties of
Carina's stars with theoretical isochrones and GCs. The conclusions in Section
\ref{sec:conclusions} close the paper. 

\section{Data Analysis}\label{sec:data}

%
\begin{figure}
\centering
\includegraphics[width=8truecm]{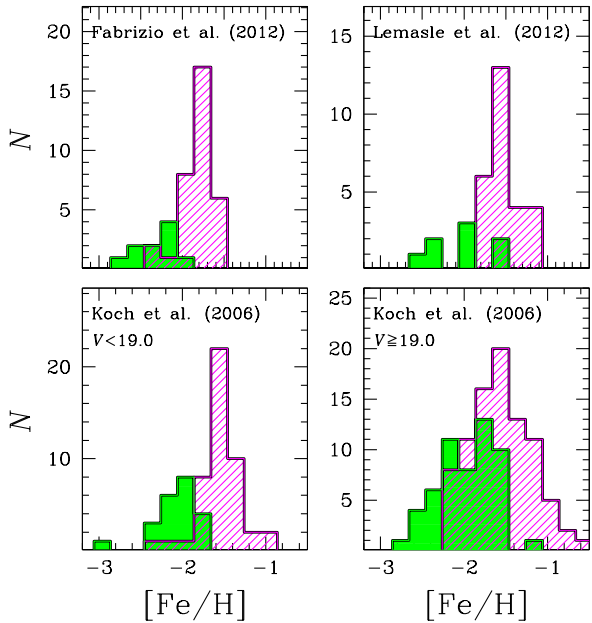}
\caption{The metallicity distribution of the three spectroscopic samples. 
The top panels show the high-resolution samples: \citet[][44 stars, left]{fabrizio12}
and \citet[][35 stars, right]{lemasle12}.
In the lower panels, the medium-resolution sample from \citet{koch06} was
split in two subgroups: stars with $V < 19$ mag (71, bottom left), and
stars with $V \geq 19$ mag (141, bottom right).
\label{fig:histo_iron}}
\end{figure}
%

	\subsection{Photometric data: splitting the Carina RGB}\label{sec:photo}

The photometric data are the same already presented in \citet{bono10}. In
particular, we focus here on the central region only, where $U$,$B$,$V$, and $I$
data overlap over an area of $\approx40\arcmin\times40\arcmin$. The left panel
of Figure \ref{fig:cubi} shows the ($V$, $B-I$) CMD of 20,240 stars selected
with the following criteria: {\em i)} they are measured in the four bands; and
{\em ii)} they are bona-fide Carina stars, with removal of most of the possible
Galactic field stars and foreground galaxies following the procedure explained
in \citet{bono10}, and based on the position of the sources in the ($U-V$),
($B-I$) colour-colour plane. The plot demonstrates a nicely cleaned CMD showing
well-defined evolutionary sequences, in particular, the two well separated
turn-offs and SGBs representing the old and intermediate--age populations,
merging into the thin RGB. The few remaining field stars populate an almost
vertical sequence at colour $1.2 < (B-I) < 1.6$ mag (cyan symbols for the
brightest ones).  These stars are virtually impossible to clean even using a
colour-colour plane, because their position is almost indistinguishable from the
Carina TO stars. Note that most of the stars bluer than the field sequence and
brighter than $V=20$ mag are Anomalous Cepheids of Carina \citep{coppola13}. The
orange crosses mark the faint part of the asymptotic giant branch (AGB) stars
sequence.

The right panel of the same figure shows the ($V$, \cubi) pseudo-CMD of
Carina. As discussed in detail by Monelli et al.\ (2013), the features in
the ($V$, \cubi) diagram are reversed compared to a standard CMD, at least in
the metallicity range covered by the Carina stars. The \cubi\ of MS stars gets
larger (i.e., less negative) as the stars get brighter, and after the turn-off
the stars evolve to lower index values along the two SGBs. In
this plane the two RGBs do not totally merge in a single, narrow sequence, as
happens in standard optical CMDs. Although some degree of overlap is present
for $-1.70 < $ \cubi\ $< -1.6$ mag, it is possible to follow the two RGB
sequences which remain well-defined, and run almost vertically and parallel
from the SGB to the tip of the RGB. In particular, the RGB stars with more
negative \cubi\ are a sequitur of the SGB of the old population, while the more
densely populated, higher \cubi\ side of the RGB is the sequitur of
the intermediate--age population. Only at the brightest magnitudes
($V\lesssim19$ mag) the two sequences diverge slightly. We stress that the
photometric error is of the order of few hundredths of magnitude along the whole
RGB, and therefore cannot account for the observed spread in the \cubi\
index. Note that the contamination due to field stars is negligible across the
RGB, while in this plane the majority of AGB stars overlie the RGB sequences.
For this reason, the latter have been removed from the following analysis.

To better characterize these two RGB sequences, we followed the procedure
outlined in Figure \ref{fig:cubi2}. Bona-fide RGB stars are shown as black open 
circles (left and central panels), while the red line, drawn by eye, is
intended to represent the right edge of their distribution. Our selection
includes only stars in the magnitude range $17.5 < V < 22$ mag. Nevertheless,
we verified that the following analysis is minimally affected, within 1$\sigma$,
by both the 
exact location of the envelope and the limit magnitude. The RGB was rectified
with respect to the red line, which is the origin of the abscissas in the
central panel. This shows, for each star, the distance in colour
$\Delta$\cubi\ from the red line. Finally, the right panel presents the
distribution of the $\Delta$\cubi\  (grey histogram). The shape of the global
distribution is best fitted by the sum (black line) of the two superimposed
gaussian curves (green and magenta lines), whose peaks are separated by
$\sim$0.08 mag. This indicates that a simple photometric index such as the
\cubi\ is able to largely separate, for the first time, the two main
components along the RGB of Carina. In particular, we conclude that stars with
\cubi\ $< -1.70$ mag),
belong to the old population, while stars with less negative index values are
predominantly intermediate--age.

We verified this finding by considering only kinematically selected stars,
adopting the sample of RGB stars from \citet{fabrizio11}. This is shown in Figure
\ref{fig:kinem}, which is a replica of Figure \ref{fig:cubi2} with superimposed
413 RGB stars with radial velocity consistent with that of Carina. By
adopting the selection from \citet{fabrizio11} and retaining stars with radial
velocities within 4$\sigma$ from the mean value, in the range +180 to +260 km 
s$^{-1}$, the occurrence of a double
peak in the pseudo-colour distribution due to the two main Carina populations
is confirmed.

	\subsection{Spectroscopic data: connecting multiple sequences and metallicity}\label{sec:spec}

 In this section, we investigate whether a connection exists between the two
sequences identified in the RGB and the chemical compositions of member stars.
To do this, we use three independent samples of spectroscopic data available in
the literature (see Figure \ref{fig:cmd_target}).  We adopt two samples of
high-resolution measurements of Fe lines, from \citet[][R$\sim$40,000, large
circles]{fabrizio12} and  \citet[][R$\sim$22,000, triangles]{lemasle12}. The two
samples include 44  and 35 stars for a total of 68 red giants (11 in common).
The entire sample  of stars with high--resolution spectra has at least one
photometric measurement in the four adopted  filters, namely $V$, and $UBI$ (see
Fig.~\ref{fig:cmd_target}). We also included red giant stars for which Fe
abundances were determined by \citet{koch06} using medium--resolution (R$\sim6000$) 
spectra (small circles in Fig.~\ref{fig:cmd_target}) in the CaT region. Out of 
the initial catalogue by \citet[][see their Tab. 6]{koch06}, including 1,197 stars, 
we selected 212 objects with [Fe/H] abundance and obying to the same photometric
and radial velocity selection criteria adopted in  Fig.  \ref{fig:cubi} and
Fig. \ref{fig:cubi2}. We note that in this analysis, we adopt the [Fe/H] abundance 
from  \citet{koch06}, and the radial velocity from \citet{fabrizio11}.

The analysis was performed splitting all the spectroscopic samples according to
their \cubi\ index. In particular, green and magenta symbols account for stars
with \cubi $< -1.70$ mag and \cubi $> -1.70$ mag, respectively. For comparison,
the right panel presents the ($V$, $B-I$) classical CMD, where the two samples
of stars selected according the \cubi\ index are mixed. Figure
\ref{fig:histo_iron} presents the metallicity distribution for the three
samples. The sample from \citet{koch06} has been further split in two,
separating the brightest (bottom left, $V < 19$ mag, 71 stars) from the
faintest (bottom right, $V \geq 19$ mag, 141 stars) ones, to account for the
change in the RGB morphology previously discussed (Figure \ref{fig:cubi}). The
plot discloses that the stars selected on the sequence with more negative \cubi\
(green histogram), identified as the RGB of the old population, are on average
more metal-poor than the stars on the other sequence (magenta), dominated by the
intermediate--age population. The different, independent samples provide a very
good overall agreement, and in particular the sample of bright stars from
\citet{koch06} notably agree with the high--resolution samples. In the case of
the faint CaT stars, the overlap between the two histogram increases, possibly
due to the increase in the observational errors. The mean metallicity of each
sample is reported in Table 1. We find that the mean metallicity
of the old population is $\approx0.5$ dex more metal-poor than the
intermediate--age one\footnote{The difference in the absolute value of the Fe
abundances between the different data sets have already been discussed in
\citet{fabrizio12}.}. Assuming the \citet{fabrizio12} data, which include stars
with both FeI and FeII measurements, we find that the mean metallicities are
[Fe/H]$=-2.32\pm0.08$ dex and [Fe/H]$=-1.82\pm0.04$ dex for the old and the
intermediate--age group, respectively. While previous investigations \citep[see
e.g.][]{lemasle12} have suggested a similar chemical evolution for Carina, we
stress that our approach is independent of isochrone fitting. The
identification of RGB stars as members of the old or the intermediate--age
population does not require previous knowledge of the metallicity.

The cut adopted to split the two quoted populations (\cubi\ = $-1.73$) mag
is,  as noted by the anonymous referee, arbitrary. Mild changes in the adopted
cut can move objects  from the old sample into the intermediate--age one and
viceversa. To constrain on a quantitative basis the impact of the adopted cut on
the mean metallicity  of the two samples, we performed the selection by adopting
\cubi\ $= -1.67 $ and \cubi\ $= -1.73$ mag. The difference was fixed according
to the photometric error  of the \cubi\ index at $V\sim$20.5 ($\sigma_{UBI}=0.03$
mag). The mean  metallicities listed in Table 1 further support
that the criterium adopted to split the two samples minimally affect the current
conclusions. In particular, the systematic difference between the mean metallicity 
of the old and the intermediate--age population is systematically, on average, 
larger than 0.5 dex.
This result is an independent demonstration that the colour spread observed in
\cubi\ is intrinsic. In fact, the \cubi\ index and the metallicity used here
have been measured in an independent way. If the RGB spread observed in \cubi\
was entirely due to photometric errors (either statistical or systematic), a
star with small (large) \cubi\ would have the same probability of being either
metal-rich or metal-poor. Moreover, this result is supported by three
independent sets of spectroscopic measurements.

%
\begin{figure}
\centering
\includegraphics[width=8truecm]{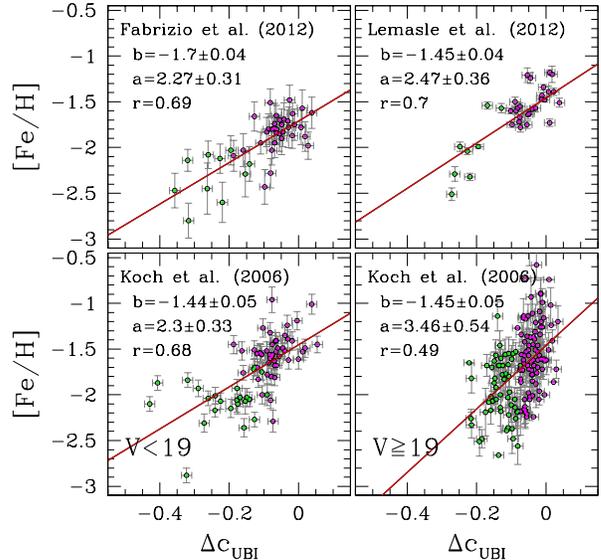}
\caption{{\em Top -} Correlation between the \cubi\ colour and the iron  content
of stars for the high--resolution samples. {\em Bottom -} The same is shown also
for the \citet{koch06} sample. The slope of the relation changes according to
the luminosity of the sample. See text for details. Green and magenta symbols
indicate stars with \cubi $< -1.70$ mag and \cubi $> -1.70$ mag, respectively,  as 
in Fig. \label{fig:cmd_target} and Fig. \label{fig:histo_iron}.
\label{fig:cubi_iron}}
\end{figure}
%

%
\begin{figure}
\centering
\includegraphics[width=8truecm]{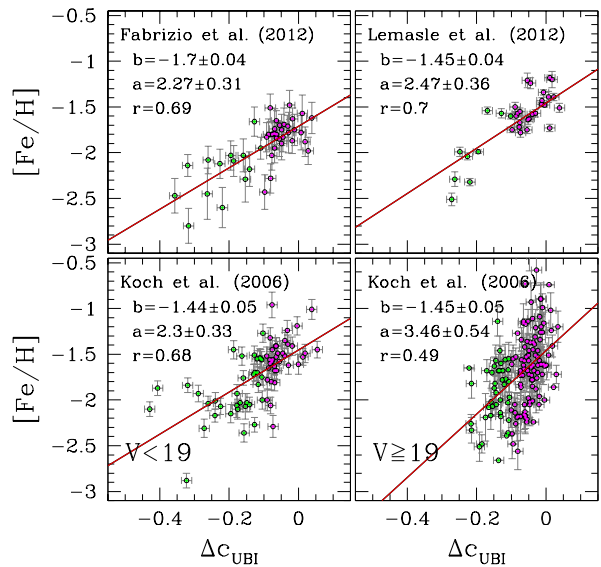}
\caption{Same as Fig. \ref{fig:cubi_iron}, but in this case green and
magenta symbols show stars with $\Delta$\cubi $< -0.1$ mag and $\Delta$\cubi 
$> -0.1$ mag, respectively. The plot demonstrates that the selection
criterium does not affect the derived conclusions.
\label{fig:cubi_iron2}}
\end{figure}
%

Figure \ref{fig:cubi_iron} presents the iron abundance as a function of the
$\Delta$\cubi\ parameter previously defined. The top panels show the trend for
the two high--resolution samples, while the lower panels present the
medium--resolution ones. The plots disclose a clear trend: the lower the
$\Delta$\cubi, the lower the Fe abundance. We fit the data with a linear function
in the form $f = ax + b$, reporting in each panel the the values of both the
zero point ($b$) and the slope ($a$), as well as the Spearman coefficient ($r$),
which suggests that the correlation is significant. The four samples provide a
general good agreement, in particular concerning the bright portion of the RGB
where both high- and medium--resolution are present. The sub-sample of faint stars
presents a steeper correlation. This can be easily explained taking into account
that the RGB morphology in the \cubi\ plane changes as a function of magnitude.
In the upper part the two RGB sequences, where most of the high--resolution stars
are, tend to diverge. This pushes the stars on the old RGB to lower \cubi,
making the relation less steep. For magnitude fainter than $V\sim19$ mag, the
two RGB are closer, and this reflects in a steeper relation between Fe and
$\Delta$\cubi. 

In \S \ref{sec:photo} we mentioned that the cut at \cubi\ $= -1.70$ mag to
split the old and the  intermediate--age is approximately equivalent to a cut at
$\Delta$\cubi\ $= -0.1$ mag.  The anonymous referee noted that the two quoted
cuts are equivalent for the high--resolution sample by \citet{lemasle12} (top
right panel of Fig.~\ref{fig:cubi_iron}) and for the faint medium--resolution
sample (bottom right panel of Fig.~\ref{fig:cubi_iron}). On the other hand,
there are objects in the bright magnitude sample and in the high-resolution
sample by \citet{fabrizio12} (left panels of Fig.~\ref{fig:cubi_iron}) moving
across the boundary of the $\Delta$\cubi distribution. To take account of the
role that the selection criterium has on the correlation between the
$\Delta$\cubi parameter and the iron abundance we performed the same  test
cutting the two populations at $\Delta$\cubi\ $= -0.1$ mag. Data plotted in Fig.
\ref{fig:cubi_iron2} show that only a few stars near the edge are classified in
a different population. Therefore, this means that we cannot reach a firm
conclusion concerning the nature (old v$_s$ intermediate--age) of the objects
that located across the boundary (\cubi $= -1.70$ mag, $\Delta$\cubi =$ -0.1$ mag) 
adopted to split the two main populations. However, the above results 
further support the evidence that the correlation is marginally dependent on
the criterium adopted to split the two populations.       

Finally, we stress that this analysis is not meant to find an absolute calibration 
of the \cubi\ index as a function of metallicity. Nevertheless, the correlation
supports the connection between the \cubi\ index and 
the Fe content of the Carina RGB stars.

%
\begin{figure}
\centering
\includegraphics[width=8truecm]{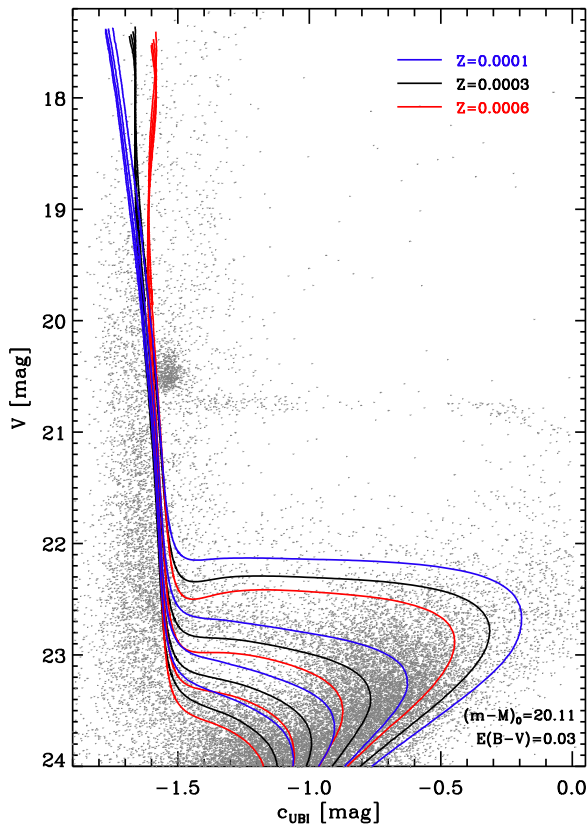}
\vspace{1cm}
\caption{Comparison with theoretical isochrones from the BaSTI database
in the (M$_{V}$, \cubi) plane. For the three metallicity labeled, isochrones
with age between 4 and 13 Gyr, in steps of 3 Gyr, are plotted. 
\label{fig:iso}}
\end{figure}
%

\section{Discussion: stellar populations and chemical evolution in Carina}\label{sec:discussion}

	\subsection{Comparison with theoretical models}
	
To assess whether the theoretical models are able to reproduce the observed
colour distribution of RGB stars, in the following we perform a comparison with
theoretical models from the
BaSTI\footnote{http://basti.oa-teramo.inaf.it/index.html} database
\citep{pietrinferni04}. Figure \ref{fig:iso} shows a comparison in the ($V$,
\cubi) plane with scaled-solar isochrones of three different metallicities
(Z=0.0001, 0.0003, 0.0006), and age ranging from 4 to 13 Gyr, in steps of 3 Gyr.
The figure discloses that the colour range covered by the selected isochrones
($\approx 0.03$ mag at $V = 21$ mag) is significantly smaller than the
observed one in the (V, \cubi) plane. Interestingly, in the upper part of the
RGB ($V \lesssim 19.5$ mag) the effect of metallicity is to split the
isochrones, in agreement with the observed increasing separation of the Carina
RGB sequences, discussed in Figure \ref{fig:cubi}. For fainter magnitudes, it is
not possible to disentangle the isochrones of different metallicity, at least in
the range typical of the Carina Fe content. The effect of age along the RGB is
largely negligible in this plane for both the old and the intermediate--age
population. Overall, we conclude that current models based on a standard
heavy-element distribution do not account for the spread in \cubi\ observed in
the Carina RGBs.

 \begin{table*}[ht!]
 \begin{center}
 \caption{Mean metallicities for the different spectroscopic samples and stellar populations}
 \begin{tabular}{lcccccc}
 \hline
 \hline
 \textrm{Spectroscopic sample}   & $<$[Fe/H]$>_{old}$ & $\sigma$  & $N_{stars}$ & $<$[Fe/H]$>_{int}$ & $\sigma$ & $N_{stars}$  \\
  \hline
\multicolumn{7}{l}{{\itshape Cut at \cubi\ = $-1.70$ mag}} \\
 \citet{fabrizio12}              & $-2.32\pm0.08$  & 0.25 &   10   &  $-1.82\pm0.03$  & 0.20 &   34    \\
 \citet{lemasle12}               & $-2.03\pm0.13$  & 0.35 &    8   &  $-1.52\pm0.03$  & 0.18 &  27    \\
 \citet{koch06} [$V < 19$ mag]   & $-2.06\pm0.05$  &  0.24 &   24  &  $-1.55\pm0.03$  &  0.24 &	47    \\
 \citet{koch06} [$V \geq 19$ mag] & $-1.96\pm0.05$  &  0.33 &   54  &  $-1.56\pm0.04$  &  0.38 &	87    \\
 \hline
\multicolumn{7}{l}{{\itshape Cut at \cubi\ = $-1.67$ mag}} \\
 \citet{fabrizio12}              & $-2.25\pm0.07$   & 0.27 & 14 &  $-1.79\pm0.03$  & 0.16  & 30   \\
 \citet{lemasle12}               & $-1.80\pm0.10$   & 0.37 & 15 &  $-1.51\pm0.04$  & 0.18  & 20   \\
 \citet{koch06} [$V < 19$ mag]   & $-1.92\pm0.06$  & 0.34 & 36 &  $-1.52\pm0.03$  & 0.19  & 35    \\
 \citet{koch06} [$V \geq 19$ mag] & $-1.85\pm0.04$  & 0.38 & 86 &  $-1.50\pm0.05$  & 0.41  & 55  \\
  \hline
\multicolumn{7}{l}{{\itshape Cut at \cubi\ = $-1.73$ mag}} \\
 \citet{fabrizio12}              & $-2.34\pm0.11$   & 0.28 & 8  &  $-1.85\pm0.04$  & 0.21  &  36  \\
 \citet{lemasle12}               & $-2.10\pm0.13$   & 0.31 & 7  &  $-1.52\pm0.03$  & 0.17  &  28  \\
 \citet{koch06} [$V < 19$ mag]   & $-2.10\pm0.06$   & 0.25 & 18 &  $-1.59\pm0.04$  & 0.26  &  53    \\
 \citet{koch06} [$V \geq 19$ mag] & $-1.93\pm0.05$   & 0.33 & 38 &  $-1.63\pm0.04$  & 0.41  &  103    \\
 \hline
 \end{tabular}
 \end{center} 
 \label{tab:tab01}
 \end{table*}

These results do not substantially change when using $\alpha$- or He-enhanced models
\citep[see also][]{cassisi13a, cassisi13b}. Current theoretical framework does not
allow us to fully explain the behaviour of the two sequences in the \cubi\ plane by
accounting for iron, age, $\alpha$-elements, and He differences alone. We remark
that this cannot be due to a residual shortcoming in the adopted colour-T$_{eff}$
transformations because we are using model predictions in a strictly differential
way.

	\subsection{Comparison with GCs}\label{sec:gc}

\begin{figure}
\centering
\includegraphics[width=9truecm, height=8truecm]{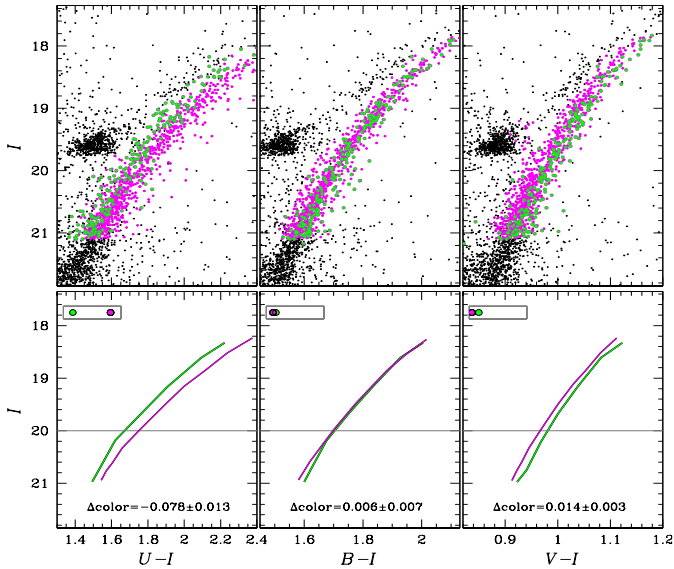}
\caption{{\em Top -} $I$ $versus$ $U-I$, $B-I$, $V-I$ CMDs zoomed in the RGB region,
with stars from the two RGB selected in the $\Delta$\cubi~ are highlighted
in green (old population) and magenta (intermediate--age). {\em Bottom -}
Ridge lines of the two populations. Note the change in the relative position 
for decreasing colour baseline. The mean colour difference with respect
of the green line at magnitude
$I=20$ mag is labeled, and illustrated by the coloured points on the top-left
corner of each panel. \label{fig:cubi_panels}}
\end{figure}

Figure \ref{fig:cubi_panels} shows how stars of the two sequences distribute
in different CMDs, namely the $U-I$, $B-I$, $V-I$ (top row, from left to
right). The two sequences appear overlapped and mixed, and therefore
indistinguishable, in $B-I$, while they are slightly segregated in $V-I$ and
substantially more so in $U-I$. However, the two sequences are swapped in
the two planes: in $U-I$ the green symbols (old stars) are 
concentrated on the blue edge of the RGB, while the magenta circles
(intermediate--age population) are located closer to the red edge. The
opposite occurs for the $V-I$ colour. This is evident in the lower
panels, where we show the fiducial lines of the two samples. To quantify the
split we calculate the colour difference at magnitude level $I$=20 mag---that
is, roughly two magnitudes below the tip of the RGB. As expected, we find
that the separation is negligible in $B-I$, and only slightly larger in
$V-I$, on the order of $0.01$ mag, which is statistically significant given the 
typical photometric error at this magnitude level. The largest separation is 
found in the $U-I$ colour, of the order of $-0.08$ mag.

%
%

It is instructive to compare the behaviour of the RGBs in Carina with
similar observations of multiple RGBs in GCs. In several GCs, the RGB and
the MS of first-population stars is redder than the second-population RGB
and MS in $B-I$, and $V-I$ colours, and the colour separation increases for
increasing colour baseline (that is, $\Delta(B-I) > \Delta(V-I) > 0$).
However, the RGBs can merge together or even change their relative
positions when using a near-UV filter ($\Delta(U-I)<0$). On the other hand,
if far-ultraviolet ($FUV$) filters are adopted the separation between the
two RGBs increases again and is significantly larger in $FUV-I$ than in
visual colours \citep[see][]{milone12a,milone12c}.

%
\begin{figure}
\centering
\includegraphics[width=8truecm]{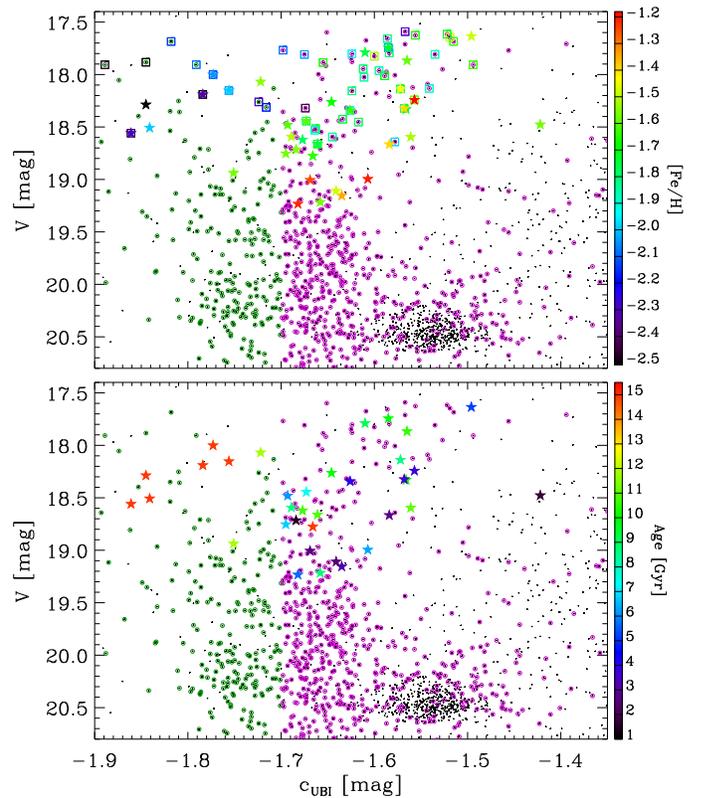}
\caption{$V$-\cubi\ diagram for bright Carina stars. The green and magenta
circles show the separation at \cubi $=-1.70$ mag performed in Fig. \ref{fig:cmd_target}.
The starred symbols highlight the 35 stars from \citet{lemasle12}.
The colour code used for these symbols, reproduced on the right of both panels,
displays the iron content (top) and age (bottom) determined by \citet{lemasle12}.
\label{fig:cubi_age}}
\end{figure}
%

%
\begin{figure}
\centering
\includegraphics[width=8truecm]{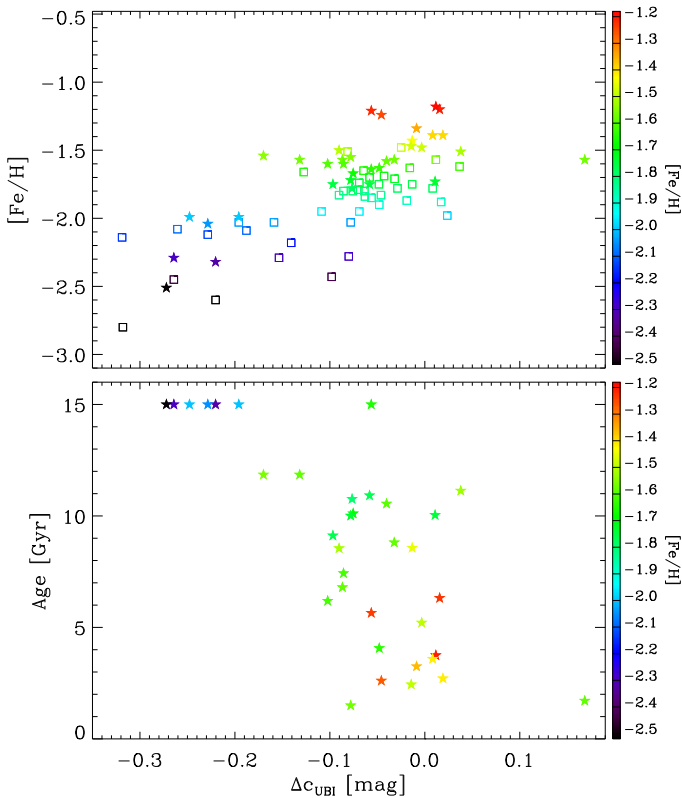}
\caption{Iron abundance (top panel) and age (bottom) from \citet{lemasle12}
as a function of the $\Delta$\cubi\ index. The colour code used in both panel
accounts for the metallicity of the 35 stars, and is the same used in the top 
panel of Fig. \ref{fig:cubi_age}. The error bars accounts for the
photometric error of our photometry and the uncertainties provided by 
\citet{lemasle12} on their age and metallicity determinations.
\label{fig:deltacubi_age}}
\end{figure}
%

This has been explained as due to the detailed abundances of the various light
elements (e.g., C, N, Na, O) and Helium. In particular, the colours of the
first-generation stars in GCs are well reproduced assuming primordial He
abundance and an O-rich/N-poor composition that is similar to that of
halo-field stars of the same metallicity. On the other hand, the colours of the
second-population stars agree with a composition in which N and Na are
enhanced---along with increased He---while C and O are depleted.
\citet{sbordone11} have demonstrated that such abundance variations mostly affect
the CN- NH- CH-bands in the UV part of the spectrum, and this translates into
different colours for the different populations when the UV filters are used.
These effects can be well reproduced by stellar models if the appropriate
chemical mixtures are adopted \citep{sbordone11}.

Results shown in Figure \ref{fig:cubi_panels} demonstrate that the RGB
behaviour of stellar populations in the Carina dSph and in GCs differ
significantly. Carina is indeed an intrinsically more complicated system
than a typical GC, due to the bursty star-formation history and the large
age and metallicity spread present in each population. Multiple populations
in a GC formed on a much shorter time-scale, and, in most cases, the internal chemical
enrichment affected only the light elements involved in the high temperature
H burning processes, while typically no spread in iron, due to Supernova
pollution, is found. 

Moving to a more speculative scenario, a possible explanation for the
empirical evidence is that light-element abundances could play a
non-negligible role in determining the behaviour of the multiple RGBs in the
Carina dSph, in analogy to what is observed in GCs. The \cubi\ index is very
sensitive to light-element variations, which are measured in dwarf galaxies
as well. However, it is not clear at present whether the chemical pattern
are similar to that of GCs. In fact, the Na-O and C-N anti-correlations have
not been observed for dSph galaxies in general, or for Carina in particular.
\citet{venn12} report measurement of one O line and also NaI
determinations for few stars, but no anti-correlation clearly stems from the
data. On the other hand, more metal-poor stars, with [Fe/H] $<-2$ dex, seem
to have lower Na content than more metal-rich ones \citep[][their Figure
12]{venn12}, in agreement with what found in GCs. Overall, C and N
abundances are yet not available. These can be crucial elements, as
the overall C+N+O abundance significantly influences the luminosity of the SGB
and hence affects the age determinations of the different populations
\citep{cassisi08,ventura09,marino12a}. Measurement of the overall C+N+O
content of intermediate--age and old stars in Carina is mandatory to properly
estimate the age of its stellar populations and thus to properly calibrate
the chemical enrichment history.

	\subsection{Comparison with previous results}

Following the referee's suggestion, we show in Fig. \ref{fig:cubi_age} a
comparison with literature values, in particular the metallicity (top panel) and
the age (bottom).  The top panel shows the $V-$\cubi\ diagram, with highlighted 
in green and magenta the two groups of stars selected in \S \ref{sec:spec}. 
The starred symbols show the 35 stars by \citet{lemasle12}, while the squares 
the 44 stars by \citet[11 objects in common][]{fabrizio12} colour-coded as labeled. 
The above data indicate that the most metal-poor stars attain the lowest \cubi\ 
values, and fall on the RGB sequence identified by the green circles. On the other 
hand, stars more metal-rich then [Fe/H]$\sim -$2, attain larger \cubi\ values, 
and all but three fall on the other sequence identified by the magenta circles.

The bottom panel of Fig.~\ref{fig:cubi_age} shows once again the two groups of stars 
selected in 
\S \ref{sec:spec} in green and in magenta together with the 35 stars by 
\citet{lemasle12}. The spectroscopic sample by \citet{fabrizio12} was not 
plotted in this panel, because the authors did not provide an individual age 
estimate. Interestingly, we found that the most metal-poor stars also have 
the largest age (saturated at 15 Gyr). This implies that all but one of these 
objects are located in the lowest \cubi\ sequence, which we identify with the 
RGB of the old population. Similarly, all the intermediate--age stars 
(t $<$ 8 Gyr) cluster in the region of the higher \cubi\ stars. On the other hand, 
starred symbols plotted with greenish colour, with estimated ages between 
$\sim$8 and $\sim$11 Gyr, attain a \cubi\ index ranging from $\sim -1.75$ to 
$\sim -1.42$ mag. In this context it is worth mentioning that the current 
comparison does not allow us to reach a firm conclusion concerning the objects 
with ages ranging from 8 to 11 Gyrs. The reasons are manifold. The uncertainty 
associated to the individual ages of RG stars is at least of the order of 
2 Gyr \citep{lemasle12}. The uncertainty on the individual iron abundance 
is at least of the order of 0.1 dex \citep{fabrizio12}. The sample is 
limited and these objects are located across the boundary adopted to split 
the old and the intermediate--age population.  

Fig. \ref{fig:deltacubi_age} shows the same [Fe/H] abundance (top) and age
(bottom) as a function of the $\Delta$\cubi\ index, with the same colour code 
adopted for metallicity in the top panel of in Fig. \ref{fig:cubi_age}. The 
above figures suggest a segregation, in the sense that stars with more negative
$\Delta$\cubi\ are systematically older and more metal-poor than those with
higher index. This further strengthens the use of \cubi\ and $\Delta$\cubi\
to separate stars from different populations. However, this result should be
cautiously treated, since other parameters together with age and iron content
are necessary to fully characterize the properties of the Carina stars (see
Fig. \ref{fig:iso} and \S\ref{sec:gc}). This means that the comparison between
isochrones and CMD along the RGB might be affected by possible systematic
effects. The anonymous referee noted that the correlation between age and
metallicity  is a thorny problem of the Carina dSph, since the two main
episodes of star formation show a well defined separation of the order of 3-4
Gyrs. According  to the recent literature, the age distribution of the two
main stellar  population in Carina  cover a broad range in age
\citep{mighell90a,smeckerhane94,smeckerhane96,mighell97,hurleykeller98,
hernandez00,dolphin02,rizzi03,io,bono10,stetson11,lemasle12}. The difference 
is mainly due to the adopted stellar isochrones  and to the diagnostic adopted
to constrain the age (MSTO v$_s$ red giants). Unfortunately, we still lack firm 
theoretical and empirical constraints to assess  whether the Carina star formation episodes were
affected by the interaction  with the Galaxy \citep{io,pasetto11,fabrizio11,lemasle12}
However, the above uncertainties do not affect our  conclusions, since we 
are only using relative ages to separate the old and  the intermediate--age 
stellar populations.  

The quoted problems do not apply to the other satellite dSphs, such as Sculptor, 
showing a well defined age--metallicity relation \citep{deboer12a,starkenburg13}. This 
stellar system appears as a fundamental benchmark to further constrain the 
use of the \cubi index to identify distinct stellar populations.

\section{Conclusions}\label{sec:conclusions}

In this paper we have investigated the stellar populations in Carina adopting
an observational approach that has been successfully used to study multiple populations in GCs.
The main results can be summarized as follows:

$\bullet$ 
The use of the \cubi\ pseudo-colour allows the discrimination
between old and intermediate--age populations. The new purely photometric
diagnostic allows us in breaking the severe age-metallicity degeneracy along a
significant portion of the RGB of this stellar system. In particular, we are
able to clearly follow the  evolutionary sequence of the old and of the
intermediate--age populations for  at least three magnitudes along the RGB.
This approach is fully supported by the analysis sample of bona-fide Carina stars
selected according to their radial velocity. \\
$\bullet$ By comparing with published spectroscopy, we found a significant
correlation between the \cubi\ index of RGB stars and their iron content. This
conclusion is supported by three independent samples of spectroscopic measurements.
We derive that the mean metallicity of the old population is on average 0.5 dex
lower than the intermediate--age one. \\
$\bullet$ A comparison with theoretical models demonstrates that current isochrones
calculated for a standard heavy-element mixture cannot reproduce the observed
distribution of stars in the ($V$, \cubi) plane.\\
$\bullet$ The appearance of Carina RGBs in different CMDs significantly differs from
what is observed for multiple populations within GCs. \\
 
The comparison between theoretical models and observations indicates that
the RGB bimodality disclosed by \cubi\ cannot be explained on the basis of a
difference in age and/or in Fe abundance only. In dealing with mono-metallic GCs, the
split of the different sub-populations is currently explained as the combination
of light-element (mainly C,N,O) and helium variations \citep[e.g.][]{marino08,
yong08, sbordone11, milone12a}. Very likely a possible difference in C, N, O
--and probably helium-- between the old and young population, combined with
metallicity and age effects, can be also responsible of the observed RGB
bimodality in Carina. However, in the case of Carina there are no measurements
for most of these elements. In addition, while in GCs C,N,O abundances are
connected with well-known correlations and anticorrelations the pattern of C,N,O
variation in this dwarf is largely unexplored \citep[e.g.][]{venn12} and the
empirical scenario is still limited by both accuracy and statistics. 

Detailed knowledge of the chemical pattern in dwarf galaxies, especially in
comparison with GCs, is very important to constrain the chemical evolution in
the Milky Way environment. The techniques shown here could be very promising
for breaking the age-metallicity degeneracy in the CMDs of stellar systems with
complex star formation history, and should be tested in different systems. In
particular, comparing Carina with other galaxies having different star 
formation histories, such as the purely old ones (Draco,Ursa Minor, Sculptor)
and those with extended star formation (Fornax, Leo~I) would be particularly
valuable.

\begin{acknowledgements}
Support for this work has been provided by the
Education and Science Ministry of Spain (grant AYA2010-16717). APM acknowledges
the  financial support from the Australian Research  Council through Discovery
Project grant DP120100475. SC is grateful for financial support from PRIN-INAF
2011 "Multiple Populations in Globular Clusters: their role in the Galaxy
assembly" (PI: E. Carretta), and from PRIN MIUR 2010-2011, project \lq{The
Chemical and Dynamical Evolution of the Milky Way and Local Group Galaxies}\rq,
prot. 2010LY5N2T (PI: F. Matteucci). MF acknowledges the financial support 
from the PO FSE Abruzzo 2007-2013 through the grant "Spectro-photometric 
characterization of stellar populations in Local Group dwarf galaxies", 
prot.89/2014/OACTe/D (PI: S. Cassisi).
\end{acknowledgements}


\begin{thebibliography}{68}
\expandafter\ifx\csname natexlab\endcsname\relax\def\natexlab#1{#1}\fi

\bibitem[{{Aparicio} {et~al.}(2001){Aparicio}, {Carrera}, \&
  {Mart{\'{\i}}nez-Delgado}}]{aparicio01}
{Aparicio}, A., {Carrera}, R., \& {Mart{\'{\i}}nez-Delgado}, D. 2001, \aj, 122,
  2524

\bibitem[{{Battaglia} {et~al.}(2012{\natexlab{a}}){Battaglia}, {Irwin},
  {Tolstoy}, {de Boer}, \& {Mateo}}]{battaglia12a}
{Battaglia}, G., {Irwin}, M., {Tolstoy}, E., {de Boer}, T., \& {Mateo}, M.
  2012{\natexlab{a}}, ArXiv e-prints

\bibitem[{{Battaglia} {et~al.}(2012{\natexlab{b}}){Battaglia}, {Irwin},
  {Tolstoy}, {de Boer}, \& {Mateo}}]{battaglia12b}
---. 2012{\natexlab{b}}, ArXiv e-prints

\bibitem[{{Bedin} {et~al.}(2004){Bedin}, {Piotto}, {Anderson}, {Cassisi},
  {King}, {Momany}, \& {Carraro}}]{bedin04}
{Bedin}, L.~R., {Piotto}, G., {Anderson}, J., {et~al.} 2004, \apjl, 605, L125

\bibitem[{{Bekki} {et~al.}(2007){Bekki}, {Campbell}, {Lattanzio}, \&
  {Norris}}]{bekki07a}
{Bekki}, K., {Campbell}, S.~W., {Lattanzio}, J.~C., \& {Norris}, J.~E. 2007,
  \mnras, 377, 335

\bibitem[{{Bellazzini} {et~al.}(2001){Bellazzini}, {Ferraro}, \&
  {Pancino}}]{bellazzini02}
{Bellazzini}, M., {Ferraro}, F.~R., \& {Pancino}, E. 2001, \mnras, 327, L15

\bibitem[{{Bono} {et~al.}(2010){Bono}, {Stetson}, {Walker}, {Monelli},
  {Fabrizio}, {Pietrinferni}, {Brocato}, {Buonanno}, {Caputo}, {Cassisi},
  {Castellani}, {Cignoni}, \& {Corsi}}]{bono10}
{Bono}, G., {Stetson}, P.~B., {Walker}, A.~R., {et~al.} 2010, \pasp, 122, 651

\bibitem[{{Cannon} {et~al.}(1998){Cannon}, {Croke}, {Bell}, {Hesser}, \&
  {Stathakis}}]{cannon98}
{Cannon}, R.~D., {Croke}, B.~F.~W., {Bell}, R.~A., {Hesser}, J.~E., \&
  {Stathakis}, R.~A. 1998, \mnras, 298, 601

\bibitem[{{Carrera} {et~al.}(2002){Carrera}, {Aparicio},
  {Mart{\'{\i}}nez-Delgado}, \& {Alonso-Garc{\'{\i}}a}}]{carrera02}
{Carrera}, R., {Aparicio}, A., {Mart{\'{\i}}nez-Delgado}, D., \&
  {Alonso-Garc{\'{\i}}a}, J. 2002, \aj, 123, 3199

\bibitem[{{Cassisi} {et~al.}(2013{\natexlab{a}}){Cassisi}, {Mucciarelli},
  {Pietrinferni}, {Salaris}, \& {Ferguson}}]{cassisi13a}
{Cassisi}, S., {Mucciarelli}, A., {Pietrinferni}, A., {Salaris}, M., \&
  {Ferguson}, J. 2013{\natexlab{a}}, \aap, 554, A19

\bibitem[{{Cassisi} {et~al.}(2013{\natexlab{b}}){Cassisi}, {Salaris}, \&
  {Pietrinferni}}]{cassisi13b}
{Cassisi}, S., {Salaris}, M., \& {Pietrinferni}, A. 2013{\natexlab{b}},
  \memsai, 84, 91

\bibitem[{{Cassisi} {et~al.}(2008){Cassisi}, {Salaris}, {Pietrinferni},
  {Piotto}, {Milone}, {Bedin}, \& {Anderson}}]{cassisi08}
{Cassisi}, S., {Salaris}, M., {Pietrinferni}, A., {et~al.} 2008, \apjl, 672,
  L115

\bibitem[{{Cole} {et~al.}(2007){Cole}, {Skillman}, {Tolstoy}, {Gallagher},
  {Aparicio}, {Dolphin}, {Gallart}, {Hidalgo}, {Saha}, {Stetson}, \&
  {Weisz}}]{cole07}
{Cole}, A.~A., {Skillman}, E.~D., {Tolstoy}, E., {et~al.} 2007, \apjl, 659, L17

\bibitem[{{Coppola} {et~al.}(2013){Coppola}, {Stetson}, {Marconi}, {Bono},
  {Ripepi}, {Fabrizio}, {Dall'Ora}, {Musella}, {Buonanno}, {Ferraro},
  {Fiorentino}, {Iannicola}, {Monelli}, {Nonino}, {Pulone}, {Th{\'e}venin}, \&
  {Walker}}]{coppola13}
{Coppola}, G., {Stetson}, P.~B., {Marconi}, M., {et~al.} 2013, \apj, 775, 6

\bibitem[{{Dall'Ora} {et~al.}(2012){Dall'Ora}, {Kinemuchi}, {Ripepi},
  {Rodgers}, {Clementini}, {Di Fabrizio}, {Smith}, {Marconi}, {Musella},
  {Greco}, {Kuehn}, {Catelan}, {Pritzl}, \& {Beers}}]{dallora12}
{Dall'Ora}, M., {Kinemuchi}, K., {Ripepi}, V., {et~al.} 2012, \apj, 752, 42

\bibitem[{{de Boer} {et~al.}(2012){de Boer}, {Tolstoy}, {Hill}, {Saha},
  {Olsen}, {Starkenburg}, {Lemasle}, {Irwin}, \& {Battaglia}}]{deboer12a}
{de Boer}, T.~J.~L., {Tolstoy}, E., {Hill}, V., {et~al.} 2012, \aap, 539, A103

\bibitem[{{D'Ercole} {et~al.}(2008){D'Ercole}, {Vesperini}, {D'Antona},
  {McMillan}, \& {Recchi}}]{dercole08}
{D'Ercole}, A., {Vesperini}, E., {D'Antona}, F., {McMillan}, S.~L.~W., \&
  {Recchi}, S. 2008, \mnras, 391, 825

\bibitem[{{Dolphin}(2002)}]{dolphin02}
{Dolphin}, A.~E. 2002, \mnras, 332, 91

\bibitem[{{Dolphin} {et~al.}(2001){Dolphin}, {Walker}, {Hodge}, {Mateo},
  {Olszewski}, {Schommer}, \& {Suntzeff}}]{dolphin01}
{Dolphin}, A.~E., {Walker}, A.~R., {Hodge}, P.~W., {et~al.} 2001, \apj, 562,
  303

\bibitem[{{Fabrizio} {et~al.}(2011){Fabrizio}, {Nonino}, {Bono}, {Ferraro},
  {Fran{\c c}ois}, {Iannicola}, {Monelli}, {Th{\'e}venin}, {Stetson}, {Walker},
  {Buonanno}, {Caputo}, {Corsi}, {Dall'Ora}, {Gilmozzi}, {James}, {Merle},
  {Pulone}, \& {Romaniello}}]{fabrizio11}
{Fabrizio}, M., {Nonino}, M., {Bono}, G., {et~al.} 2011, \pasp, 123, 384

\bibitem[{{Fabrizio} {et~al.}(2012){Fabrizio}, {Merle}, {Th{\'e}venin},
  {Nonino}, {Bono}, {Stetson}, {Ferraro}, {Iannicola}, {Monelli}, {Walker},
  {Buonanno}, {Caputo}, {Corsi}, {Dall''Ora}, {Degl''Innocenti}, {Fran{\c
  c}ois}, {Gilmozzi}, {Marconi}, {Pietrinferni}, {Prada Moroni}, {Primas},
  {Pulone}, {Ripepi}, \& {Romaniello}}]{fabrizio12}
{Fabrizio}, M., {Merle}, T., {Th{\'e}venin}, F., {et~al.} 2012, \pasp, 124, 519

\bibitem[{{Gallagher} {et~al.}(1998){Gallagher}, {Tolstoy}, {Dohm-Palmer},
  {Skillman}, {Cole}, {Hoessel}, {Saha}, \& {Mateo}}]{gallagher98}
{Gallagher}, J.~S., {Tolstoy}, E., {Dohm-Palmer}, R.~C., {et~al.} 1998, \aj,
  115, 1869

\bibitem[{{Gallart} {et~al.}(1996){Gallart}, {Aparicio}, \&
  {Vilchez}}]{gallart96a}
{Gallart}, C., {Aparicio}, A., \& {Vilchez}, J.~M. 1996, \aj, 112, 1928

\bibitem[{{Grundahl} {et~al.}(1999){Grundahl}, {Catelan}, {Landsman},
  {Stetson}, \& {Andersen}}]{grundahl99}
{Grundahl}, F., {Catelan}, M., {Landsman}, W.~B., {Stetson}, P.~B., \&
  {Andersen}, M.~I. 1999, \apj, 524, 242

\bibitem[{{Held} {et~al.}(1999){Held}, {Saviane}, \& {Momany}}]{held99}
{Held}, E.~V., {Saviane}, I., \& {Momany}, Y. 1999, \aap, 345, 747

\bibitem[{{Helmi} {et~al.}(2006){Helmi}, {Irwin}, {Tolstoy}, {Battaglia},
  {Hill}, {Jablonka}, {Venn}, {Shetrone}, {Letarte}, {Arimoto}, {Abel},
  {Francois}, {Kaufer}, {Primas}, {Sadakane}, \& {Szeifert}}]{helmi06}
{Helmi}, A., {Irwin}, M.~J., {Tolstoy}, E., {et~al.} 2006, \apjl, 651, L121

\bibitem[{{Hernandez} {et~al.}(2000){Hernandez}, {Gilmore}, \&
  {Valls-Gabaud}}]{hernandez00}
{Hernandez}, X., {Gilmore}, G., \& {Valls-Gabaud}, D. 2000, \mnras, 317, 831

\bibitem[{{Hidalgo} {et~al.}(2011){Hidalgo}, {Aparicio}, {Skillman}, {Monelli},
  {Gallart}, {Cole}, {Dolphin}, {Weisz}, {Bernard}, {Cassisi}, {Mayer},
  {Stetson}, {Tolstoy}, \& {Ferguson}}]{hidalgo11}
{Hidalgo}, S.~L., {Aparicio}, A., {Skillman}, E., {et~al.} 2011, \apj, 730, 14

\bibitem[{{Hurley-Keller} {et~al.}(1998){Hurley-Keller}, {Mateo}, \&
  {Nemec}}]{hurleykeller98}
{Hurley-Keller}, D., {Mateo}, M., \& {Nemec}, J. 1998, \aj, 115, 1840

\bibitem[{{Koch} {et~al.}(2006){Koch}, {Grebel}, {Wyse}, {Kleyna}, {Wilkinson},
  {Harbeck}, {Gilmore}, \& {Evans}}]{koch06}
{Koch}, A., {Grebel}, E.~K., {Wyse}, R.~F.~G., {et~al.} 2006, \aj, 131, 895

\bibitem[{{Kraft} {et~al.}(1992){Kraft}, {Sneden}, {Langer}, \&
  {Prosser}}]{kraft92}
{Kraft}, R.~P., {Sneden}, C., {Langer}, G.~E., \& {Prosser}, C.~F. 1992, \aj,
  104, 645

\bibitem[{{Lee} {et~al.}(2009){Lee}, {Kang}, {Lee}, \& {Lee}}]{lee09}
{Lee}, J.-W., {Kang}, Y.-W., {Lee}, J., \& {Lee}, Y.-W. 2009, \nat, 462, 480

\bibitem[{{Lemasle} {et~al.}(2012){Lemasle}, {Hill}, {Tolstoy}, {Venn},
  {Shetrone}, {Irwin}, {de Boer}, {Starkenburg}, \& {Salvadori}}]{lemasle12}
{Lemasle}, B., {Hill}, V., {Tolstoy}, E., {et~al.} 2012, \aap, 538, A100

\bibitem[{{Marino} {et~al.}(2008){Marino}, {Villanova}, {Piotto}, {Milone},
  {Momany}, {Bedin}, \& {Medling}}]{marino08}
{Marino}, A.~F., {Villanova}, S., {Piotto}, G., {et~al.} 2008, \aap, 490, 625

\bibitem[{{Marino} {et~al.}(2012){Marino}, {Milone}, {Sneden}, {Bergemann},
  {Kraft}, {Wallerstein}, {Cassisi}, {Aparicio}, {Asplund}, {Bedin}, {Hilker},
  {Lind}, {Momany}, {Piotto}, {Roederer}, {Stetson}, \& {Zoccali}}]{marino12a}
{Marino}, A.~F., {Milone}, A.~P., {Sneden}, C., {et~al.} 2012, \aap, 541, A15

\bibitem[{{Martin} {et~al.}(2009){Martin}, {McConnachie}, {Irwin}, {Widrow},
  {Ferguson}, {Ibata}, {Dubinski}, {Babul}, {Chapman}, {Fardal}, {Lewis},
  {Navarro}, \& {Rich}}]{martin09}
{Martin}, N.~F., {McConnachie}, A.~W., {Irwin}, M., {et~al.} 2009, \apj, 705,
  758

\bibitem[{{McConnachie} \& {Irwin}(2006)}]{mcconnachie06}
{McConnachie}, A.~W., \& {Irwin}, M.~J. 2006, \mnras, 365, 1263

\bibitem[{{Mighell}(1990)}]{mighell90a}
{Mighell}, K.~J. 1990, \aaps, 82, 1

\bibitem[{{Mighell}(1997)}]{mighell97}
---. 1997, \aj, 114, 1458

\bibitem[{{Milone} {et~al.}(2012{\natexlab{a}}){Milone}, {Marino}, {Piotto},
  {Bedin}, {Anderson}, {Aparicio}, {Cassisi}, \& {Rich}}]{milone12c}
{Milone}, A.~P., {Marino}, A.~F., {Piotto}, G., {et~al.} 2012{\natexlab{a}},
  \apj, 745, 27

\bibitem[{{Milone} {et~al.}(2012{\natexlab{b}}){Milone}, {Piotto}, {Bedin},
  {Cassisi}, {Anderson}, {Marino}, {Pietrinferni}, \& {Aparicio}}]{milone12e}
{Milone}, A.~P., {Piotto}, G., {Bedin}, L.~R., {et~al.} 2012{\natexlab{b}},
  \aap, 537, A77

\bibitem[{{Milone} {et~al.}(2008){Milone}, {Bedin}, {Piotto}, {Anderson},
  {King}, {Sarajedini}, {Dotter}, {Chaboyer}, {Mar{\'{\i}}n-Franch},
  {Majewski}, {Aparicio}, {Hempel}, {Paust}, {Reid}, {Rosenberg}, \&
  {Siegel}}]{milone08}
{Milone}, A.~P., {Bedin}, L.~R., {Piotto}, G., {et~al.} 2008, \apj, 673, 241

\bibitem[{{Milone} {et~al.}(2012{\natexlab{c}}){Milone}, {Marino}, {Cassisi},
  {Piotto}, {Bedin}, {Anderson}, {Allard}, {Aparicio}, {Bellini}, {Buonanno},
  {Monelli}, \& {Pietrinferni}}]{milone12a}
{Milone}, A.~P., {Marino}, A.~F., {Cassisi}, S., {et~al.} 2012{\natexlab{c}},
  \apjl, 754, L34

\bibitem[{{Momany} {et~al.}(2005){Momany}, {Held}, {Saviane}, {Bedin},
  {Gullieuszik}, {Clemens}, {Rizzi}, {Rich}, \& {Kuijken}}]{momany05}
{Momany}, Y., {Held}, E.~V., {Saviane}, I., {et~al.} 2005, \aap, 439, 111

\bibitem[{{Monelli} {et~al.}(2010{\natexlab{a}}){Monelli}, {Cassisi},
  {Bernard}, {Hidalgo}, {Aparicio}, {Gallart}, \& {Skillman}}]{monelli10a}
{Monelli}, M., {Cassisi}, S., {Bernard}, E.~J., {et~al.} 2010{\natexlab{a}},
  \apj, 718, 707

\bibitem[{{Monelli} {et~al.}(2003){Monelli}, {Pulone}, {Corsi}, {Castellani},
  {Bono}, {Walker}, {Brocato}, {Buonanno}, {Caputo}, {Castellani}, {Dall'Ora},
  {Marconi}, {Nonino}, {Ripepi}, \& {Smith}}]{io}
{Monelli}, M., {Pulone}, L., {Corsi}, C.~E., {et~al.} 2003, \aj, 126, 218

\bibitem[{{Monelli} {et~al.}(2010{\natexlab{b}}){Monelli}, {Hidalgo},
  {Stetson}, {Aparicio}, {Gallart}, {Dolphin}, {Cole}, {Weisz}, {Skillman},
  {Bernard}, {Mayer}, {Navarro}, {Cassisi}, {Drozdovsky}, \&
  {Tolstoy}}]{monelli10b}
{Monelli}, M., {Hidalgo}, S.~L., {Stetson}, P.~B., {et~al.} 2010{\natexlab{b}},
  \apj, 720, 1225

\bibitem[{{Monelli} {et~al.}(2013){Monelli}, {Milone}, {Stetson}, {Marino},
  {Cassisi}, {del Pino Molina}, {Salaris}, {Aparicio}, {Asplund}, {Grundahl},
  {Piotto}, {Weiss}, {Carrera}, {Cebri{\'a}n}, {Murabito}, {Pietrinferni}, \&
  {Sbordone}}]{monelli13}
{Monelli}, M., {Milone}, A.~P., {Stetson}, P.~B., {et~al.} 2013, \mnras

\bibitem[{{Okamoto} {et~al.}(2012){Okamoto}, {Arimoto}, {Yamada}, \&
  {Onodera}}]{okamoto12}
{Okamoto}, S., {Arimoto}, N., {Yamada}, Y., \& {Onodera}, M. 2012, \apj, 744,
  96

\bibitem[{{Pasetto} {et~al.}(2011){Pasetto}, {Grebel}, {Berczik}, {Chiosi}, \&
  {Spurzem}}]{pasetto11}
{Pasetto}, S., {Grebel}, E.~K., {Berczik}, P., {Chiosi}, C., \& {Spurzem}, R.
  2011, \aap, 525, A99

\bibitem[{{Pietrinferni} {et~al.}(2004){Pietrinferni}, {Cassisi}, {Salaris}, \&
  {Castelli}}]{pietrinferni04}
{Pietrinferni}, A., {Cassisi}, S., {Salaris}, M., \& {Castelli}, F. 2004, \apj,
  612, 168

\bibitem[{{Piotto} {et~al.}(2012){Piotto}, {Milone}, {Anderson}, {Bedin},
  {Bellini}, {Cassisi}, {Marino}, {Aparicio}, \& {Nascimbeni}}]{piotto12}
{Piotto}, G., {Milone}, A.~P., {Anderson}, J., {et~al.} 2012, \apj, 760, 39

\bibitem[{{Rizzi} {et~al.}(2003){Rizzi}, {Held}, {Bertelli}, \&
  {Saviane}}]{rizzi03}
{Rizzi}, L., {Held}, E.~V., {Bertelli}, G., \& {Saviane}, I. 2003, \apjl, 589,
  L85

\bibitem[{{Rizzi} {et~al.}(2007){Rizzi}, {Held}, {Saviane}, {Tully}, \&
  {Gullieuszik}}]{rizzi07}
{Rizzi}, L., {Held}, E.~V., {Saviane}, I., {Tully}, R.~B., \& {Gullieuszik}, M.
  2007, \mnras, 380, 1255

\bibitem[{{Sanna} {et~al.}(2010){Sanna}, {Bono}, {Stetson}, {Ferraro},
  {Monelli}, {Nonino}, {Prada Moroni}, {Bresolin}, {Buonanno}, {Caputo},
  {Cignoni}, {Degl'Innocenti}, {Iannicola}, {Matsunaga}, {Pietrinferni},
  {Romaniello}, {Storm}, \& {Walker}}]{sanna10}
{Sanna}, N., {Bono}, G., {Stetson}, P.~B., {et~al.} 2010, \apjl, 722, L244

\bibitem[{{Sbordone} {et~al.}(2011){Sbordone}, {Salaris}, {Weiss}, \&
  {Cassisi}}]{sbordone11}
{Sbordone}, L., {Salaris}, M., {Weiss}, A., \& {Cassisi}, S. 2011, \aap, 534,
  A9

\bibitem[{{Shetrone} {et~al.}(2003){Shetrone}, {Venn}, {Tolstoy}, {Primas},
  {Hill}, \& {Kaufer}}]{shetrone03}
{Shetrone}, M., {Venn}, K.~A., {Tolstoy}, E., {et~al.} 2003, \aj, 125, 684

\bibitem[{{Shetrone} {et~al.}(2001){Shetrone}, {C{\^o}t{\'e}}, \&
  {Sargent}}]{shetrone01}
{Shetrone}, M.~D., {C{\^o}t{\'e}}, P., \& {Sargent}, W.~L.~W. 2001, \apj, 548,
  592

\bibitem[{{Skillman} {et~al.}(2003){Skillman}, {Tolstoy}, {Cole}, {Dolphin},
  {Saha}, {Gallagher}, {Dohm-Palmer}, \& {Mateo}}]{skillman03}
{Skillman}, E.~D., {Tolstoy}, E., {Cole}, A.~A., {et~al.} 2003, \apj, 596, 253

\bibitem[{{Smecker-Hane} {et~al.}(1994){Smecker-Hane}, {Stetson}, {Hesser}, \&
  {Lehnert}}]{smeckerhane94}
{Smecker-Hane}, T.~A., {Stetson}, P.~B., {Hesser}, J.~E., \& {Lehnert}, M.~D.
  1994, \aj, 108, 507

\bibitem[{{Smecker-Hane} {et~al.}(1996){Smecker-Hane}, {Stetson}, {Hesser}, \&
  {Vandenberg}}]{smeckerhane96}
{Smecker-Hane}, T.~A., {Stetson}, P.~B., {Hesser}, J.~E., \& {Vandenberg},
  D.~A. 1996, in Astronomical Society of the Pacific Conference Series,
  Vol.~98, From Stars to Galaxies: the Impact of Stellar Physics on Galaxy
  Evolution, ed. C.~{Leitherer}, U.~{Fritze-von-Alvensleben}, \& J.~{Huchra},
  328

\bibitem[{{Starkenburg} {et~al.}(2013){Starkenburg}, {Hill}, {Tolstoy},
  {Fran{\c c}ois}, {Irwin}, {Boschman}, {Venn}, {de Boer}, {Lemasle},
  {Jablonka}, {Battaglia}, {Groot}, \& {Kaper}}]{starkenburg13}
{Starkenburg}, E., {Hill}, V., {Tolstoy}, E., {et~al.} 2013, \aap, 549, A88

\bibitem[{{Stetson} {et~al.}(2011){Stetson}, {Monelli}, {Fabrizio}, {Walker},
  {Bono}, {Buonanno}, {Caputo}, {Cassisi}, {Corsi}, {Dall'Ora},
  {Degl'Innocenti}, {Fran{\c c}ois}, {Ferraro}, {Gilmozzi}, {Iannicola},
  {Merle}, {Nonino}, {Pietrinferni}, {Moroni}, {Pulone}, {Romaniello}, \&
  {Th{\'e}venin}}]{stetson11}
{Stetson}, P.~B., {Monelli}, M., {Fabrizio}, M., {et~al.} 2011, The Messenger,
  144, 32

\bibitem[{{Tolstoy} {et~al.}(2003){Tolstoy}, {Venn}, {Shetrone}, {Primas},
  {Hill}, {Kaufer}, \& {Szeifert}}]{tolstoy03}
{Tolstoy}, E., {Venn}, K.~A., {Shetrone}, M., {et~al.} 2003, \aj, 125, 707

\bibitem[{{Venn} {et~al.}(2012){Venn}, {Shetrone}, {Irwin}, {Hill}, {Jablonka},
  {Tolstoy}, {Lemasle}, {Divell}, {Starkenburg}, {Letarte}, {Baldner},
  {Battaglia}, {Helmi}, {Kaufer}, \& {Primas}}]{venn12}
{Venn}, K.~A., {Shetrone}, M.~D., {Irwin}, M.~J., {et~al.} 2012, \apj, 751, 102

\bibitem[{{Ventura} {et~al.}(2009){Ventura}, {Caloi}, {D'Antona}, {Ferguson},
  {Milone}, \& {Piotto}}]{ventura09}
{Ventura}, P., {Caloi}, V., {D'Antona}, F., {et~al.} 2009, \mnras, 399, 934

\bibitem[{{Walker} {et~al.}(2009){Walker}, {Mateo}, \& {Olszewski}}]{walker09}
{Walker}, M.~G., {Mateo}, M., \& {Olszewski}, E.~W. 2009, \aj, 137, 3100

\bibitem[{{Yong} \& {Grundahl}(2008)}]{yong08}
{Yong}, D., \& {Grundahl}, F. 2008, \apjl, 672, L29

\end{thebibliography}

\end{document}